\documentclass[aps,prc,preprint,superscriptaddress,showpacs,showkeys]{revtex4}
\usepackage{graphicx}
\usepackage{amsfonts,amsmath,amssymb,bm}

\begin{document}

\title{Divergence-type $2+1$ dissipative hydrodynamics applied to heavy-ion collisions}

\author{J. Peralta-Ramos}
\email{jperalta@df.uba.ar}

\affiliation{Departamento de F\'isica, FCEyN, Universidad de Buenos Aires \\
Instituto de F\'isica de Buenos Aires, CONICET \\
Ciudad Universitaria, Pabell\'on I, 1428 Buenos Aires, Argentina}

\author{E. Calzetta}
\email{calzetta@df.uba.ar}
\affiliation{Departamento de F\'isica, FCEyN, Universidad de Buenos Aires \\
Instituto de F\'isica de Buenos Aires, CONICET \\
Ciudad Universitaria, Pabell\'on I, 1428 Buenos Aires, Argentina}

\date{\today}

\begin{abstract}
\end{abstract}


\begin{abstract}
We apply divergence-type theory (DTT) dissipative hydrodynamics to study the $2+1$ space-time evolution of the fireball created in Au+Au relativistic heavy-ion collisions at $\sqrt{s_{NN}}=$200 GeV. DTTs are exact hydrodynamic theories that do no rely on velocity gradient expansions and therefore go beyond second-order theories. We numerically solve the equations of motion of the DTT for Glauber initial conditions and compare the results with those of second-order theory based on conformal invariance (BRSS) and with data. We find that the charged-hadron minumum-bias elliptic flow reaches its maximum value at lower $p_T$ in the DTT, and that the DTT allows for a value of $\eta/s$ slightly larger than that of the BRSS. Our results show that the differences between viscous hydrodynamic formalisms are a significant source of uncertainty in the precise extraction of $\eta/s$ from experiments. 
\end{abstract}

\maketitle

%\tableofcontents

\section{Introduction}

The heavy ion collisions experiments performed at BNL’s Relativistic Heavy Ion Collider (RHIC)
create a hot and dense medium, the Quark-Gluon plasma (QGP). One of the most important discoveries at RHIC is the large elliptic flow in non-central Au+Au collisions, which is a clear indication of collective behavior. By now, it is generally agreed that the QGP thermalizes on times $\lesssim$ 2.5 fm/c and behaves as a fluid with one of the lowest viscosity-to-entropy ratio observed in nature $\eta/s \lesssim $ 0.5 \cite{phenixwhite,bound,lac}.

In recent years, relativistic hydrodynamics has become an efficient tool for describing the evolution of the fireball created at RHIC (there is a vast literature on this subject, see for instance Refs. \cite{rom09,libro,muronga04,houv,hsc06,ris,heinz09,ads,chaud,phenixwhite,teanrev,soren,shrev}). Ideal hydrodynamics has been partly successful in explaining the observed collective flow at low transverse momentum and in central collisions \cite{ideal}. Nevertheless, one should notice that when using a realistic equation of state (including a crossover phase transition) and allowing for separate kinetic and chemical freeze outs, it seems difficult to fit the data with ideal hydrodynamics \cite{phenixwhite,star2010,huovani,songthesis,luzum,luzum2}. 
Moreover, if one aims eventually to derive the QGP viscosity from experimental data, one must start from a theoretical
framework which allows for such effects. 

When one attempts to formulate a relativistic real hydrodynamics one finds that there is simply no equivalent to the non-relativistic Navier-Stokes equations. A straightforward relativistic generalization of the Navier-Stokes equations yields the so-called first order theories. These theories are plagued with causality and stability problems \cite{rom09,libro,israel}. One is therefore led to consider the so-called second-order theories (SOTs). These theories are presented as an expansion of the viscous tensor in velocity gradients, neglecting all orders higher than the second. They are  unreliable in situations where these gradients are strong, and indeed they are known to fail, for example, in the description of strong shocks \cite{shock,jou}. It is then valuable to develop alernative theories, not limited to weak velocity gradients, to provide at least an estimate of the expected accuracy of the gradient expansion.

With this in mind, in Ref. \cite{nos} the present authors developed an hydrodynamical description of a conformal field within the framework of the so-called divergence-type theories (DTT) developed by Geroch \cite{geroch} (see also Refs. \cite{geraof,liu,calz98}). DTTs do not rely on velocity gradient expansions and in this sense they go beyond second-order theories. The purpose of this work is to present numerical results obtained from solving the equations of the DTT in $2+1$ dimensions. We use the equations to simulate Au+Au collisions, and compare the results both to experimental data and to a representative SOT. As was done in previous studies by other groups, we limit ourselves to boost-invariant longitudinal expansion in flat space-time. 

In the last years, there have been numerous theoretical studies of viscous hydrodynamics in $2+1$ dimensions applied to heavy ion collisions \cite{luzum,luzum2,dus08,song08,chaud,romyrom,hsc06,mota}. Most previous works employ Israel-Stewart formalism or some variation of it in order to simulate the evolution of the fireball. Very recently, Luzum and Romatschke \cite{luzum,luzum2} perfomed detailed simulations of Au+Au collisions based on the conformal hydrodynamical equations. The consistent picture that emerges from these diverse studies is that it is possible to match viscous hydrodynamics results to experimental data, provided $\eta/s \lesssim $ 5-6 $\times$ 1/4$\pi$ \cite{luzum,luzum2,bound,heinz09}. See also \cite{ael,lub}.

In this paper we shall choose as prototype SOT the one developed by Baier et al \cite{sonhydro} (see also Bhattacharyya et al \cite{bat} and Natsuume et al \cite{nat}). For simplicity we will refer to this hydrodynamic theory as BRSS (Baier-Romatschke-Son-Starinets). The BRSS theory is based on conformal invariance and extends the well-known Israel-Stewart (IS) formalism \cite{israel} in that it contains {\it all} second-order terms that can appear in the stress-energy tensor of a conformal fluid \cite{rom09,kin}. The reason to study conformal field hydrodynamics is that it is relevant to the QGP since, as shown by Lattice calculations \cite{panero}, QCD is approximately conformal at high temperatures. In addition, a wealth of information of the strongly-interacting plasma such as transport coefficients (inaccesible to kinetic theory) can be obtained from the AdS/CFT correspondence \cite{ads,sonhydro,bat,nat}.

The main results we arrive at are: (i) the momentum anisotropy is smaller in the DTT than in the BRSS, (ii) the charged-hadron minumum-bias elliptic flow reaches its maximum value at lower $p_T$ in the DTT, and (iii) the matching of DTT results to data allows a viscosity-to-entropy ratio slightly larger than that of the BRSS. 

We should note here that in any hydrodynamic simulation there are numerous sources of uncertainty, such as those coming from the initial conditions, from the freeze-out procedure and from hadron dynamics (just to mention a few), which unfortunately prevent a precise determination of $\eta/s$. Ours is not the exception, and here we merely intend to show that the DTT is an alternative to SOTs in the modeling of heavy-ion collisions (an alternative which may prove useful in those cases where large velocity gradients are present - e.g. shock-waves \cite{shock}). The results we obtain also show that the differences between hydrodynamic formalisms are a significant source of uncertainty in the precise extraction of $\eta/s$ from data. This indubitably points to the conclusion that, until these uncertainties are under control, care should be taken when attempting to extract $\eta/s$ from hydrodynamic simulations. In this sense, the values for $\eta/s$ presented in this work should be regarded as rough estimates. 

The paper is organized as follows. In section \ref{theo} we present the $2+1$ hydrodynamic equations of the BRSS and of the DTT, and describe the initial conditions, the equation of state (EoS) and the freeze-out prescription employed in the simulations. In section \ref{res} we present and discuss the results obtained, and in section \ref{concsec} we present our conclusions. In Appendix \ref{dep} we evaluate the sensitivity of the results on the values of second-order transport coefficients and on the spatial mesh used in the simulations. In Appendix \ref{dtt} we give a brief overview of divergence-type theories and derive Eq. (\ref{dxi}). 

\section{Theoretical setup}
\label{theo}

\subsection{Hydrodynamic equations}
\label{hydro}

In this section we present the hydrodynamic equations of the BRSS and of the DTT for boost-invariant flow in $2+1$ dimensions. We employ Milne coordinates defined by proper time $\tau=\sqrt{t^2-z^2}$ and rapidity $\psi=\textrm{arctanh}(z/t)$, and, as mentioned in the Introduction, work in flat space-time. In these coordinates the metric tensor reads $g_{\mu\nu}=(1,-1,-1,-\tau^2)$. It is convenient to use Cartesian coordinates $(x,y)$ in the transverse plane (instead of polar coordinates) since in this way the only non-vanishing Christoffel symbols are $\Gamma^\tau_{\psi\psi}=\tau$ and $\Gamma^\psi_{\tau\psi}=1/\tau$. The fluid velocity is $\vec{u}=(u^\tau,u^x,u^y,0)$ and is normalized as $u_\mu u^\mu=1$. 

The stress-energy tensor for dissipative relativistic hydrodynamics is 
\begin{equation}
\begin{split}
T^{\mu\nu} &=\rho u^\mu u^\nu - p \Delta^{\mu\nu} + \Pi^{\mu\nu} ~~~ \textrm{with} \\
\Delta^{\mu\nu} &= g^{\mu\nu}-u^\mu u^\nu
\end{split}
\end{equation}
where $\rho$ and $p$ are the energy density and the pressure in the local rest frame, and $\Pi^{\mu\nu}$ is the viscous shear tensor which is transverse ($u_\mu \Pi^{\mu\nu}=0$), traceless and symmetric. The tensor $\Delta^{\mu\nu}$ is the spatial projector orthogonal to $u^\mu$. For a conformal fluid we have $T^\mu_\mu=0$, so $\rho=3p$ and the bulk viscosity vanishes. 

In what follows, Latin indices stand for transverse coordinates $(x,y)$, $D^\mu$ is the geometric covariant derivative, $D=u_\mu D^\mu$ and $\nabla^\mu=\Delta^{\mu\nu}D_\nu$ are the comoving time and space derivatives, respectively, and $<\ldots>$ denote the spatial, symmetric and traceless projection of a tensor:
\begin{equation}
A^{<\mu\nu>}= (\frac{1}{2}\Delta^{\mu\alpha}\Delta^{\gamma\nu}+\frac{1}{2}\Delta^{\mu\gamma}\Delta^{\alpha\nu} -\frac{1}{3}\Delta^{\mu\nu}\Delta^{\alpha\gamma})A_{\alpha\gamma} ~.
\end{equation}

The hydrodynamic equations are the conservation equations for the stress-energy tensor together with the evolution equation for the shear tensor $\Pi^{\mu\nu}$. The former reads:
 
\begin{equation}
\begin{split}
(\rho+p)Du^i &= \frac{1}{3}(g^{ij}\partial_j \rho - u^i u^\alpha \partial_\alpha \rho) - \Delta^i_\alpha D_\beta \Pi^{\alpha\beta} \\
D\rho &= -(\rho+p)\nabla_\mu u^\mu + \Pi^{\mu\nu}\sigma_{\mu\nu}
\end{split}
\label{conseq}
\end{equation}
where

\begin{equation}
\begin{split}
D_\beta \Pi^{\alpha\beta} &= \Pi^{i\alpha}\partial_\tau \frac{u_i}{u_\tau} + \frac{u_i}{u_\tau}\partial_\tau\Pi^{i\alpha} + \partial_i \Pi^{i\alpha} \\
&~ + \Gamma^\alpha_{\beta\gamma}\Pi^{\beta \gamma} + \Gamma^\beta_{\beta\gamma}\Pi^{\alpha\gamma} ~.
\end{split}
\label{dpiformal}
\end{equation}

In the BRSS, the evolution of the shear tensor is given by \cite{sonhydro}
\begin{equation}
\begin{split}
\partial_\tau \Pi^{i\alpha} &= -\frac{4}{3u^\tau}\Pi^{i\alpha}\nabla_\mu u^\mu - \frac{1}{\tau_\pi u^\tau}\Pi^{i\alpha} + \frac{\eta}{\tau_\pi u^\tau} \sigma^{i\alpha} \\
&~ - \frac{\lambda_1}{2\tau_\pi \eta^2 u^\tau}\Pi^{<i}_\mu \Pi^{\alpha> \mu} - \frac{u^i\Pi^\alpha_\mu + u^\alpha \Pi^i_\mu}{u^\tau}Du^\mu \\
&~ -\frac{u^j}{u^\tau}\partial_j \Pi^{i\alpha} 
\end{split}
\label{dpi}
\end{equation}
where $\eta$ is the shear viscosity, $(\tau_\pi,\lambda_1)$ are second-order transport coefficients, 
\begin{equation}
\sigma^{\mu\nu}=\nabla^{<\mu}u^{\nu>}
\end{equation}
is the first-order shear tensor, and 
\begin{equation}
\begin{split}
\nabla_\mu u^\mu &= \partial_\tau u^\tau + \partial_i u^i + \frac{u^\tau}{\tau} \\
\nabla_{<x}u_{x>} &= \Delta^{\tau x}\partial_\tau u^x + \Delta^{ix}\partial_i u^x - \frac{1}{3}\Delta^{xx} \nabla_\mu u^\mu \\
\nabla_{<x} u_{y>} &= \frac{1}{2}\Delta^{\tau x}\partial_\tau u^y + \frac{1}{2}\Delta^{\tau y}\partial_\tau u^x +
\frac{1}{2}\Delta^{ix}\partial_i u^y\\
&~ +\frac{1}{2}\Delta^{iy}\partial_i u^x
- \frac{1}{3}\Delta^{xy} \nabla_\mu u^\mu \\
\nabla_{<\psi}u_{\psi>} &= \tau^4\Delta^{\psi\psi} \Gamma^\psi_{\tau\psi} u^\tau -\frac{1}{3}\tau^4 \Delta^{\psi\psi}\nabla_\mu u^\mu ~.
\end{split}
\label{extras1}
\end{equation}

We note that in Eq. (\ref{dpi}) we have neglected terms involving the fluid vorticity, which would be multiplied by additional second-order transport coefficients $\lambda_{2}$ and $\lambda_{3}$ \cite{sonhydro,bat}. The reason is that, for two dimensional flow, it can be shown (see Ref. \cite{luzum,luzum2} and references therein) that if the vorticity is zero initially (as it is the case here), it will remain negligible throughout the evolution up to terms which are third-order in velocity gradients (therefore beyond the scope of second-order theory). As already mentioned in the Introduction, the IS formalism is contained in the BRSS equations, as can be seen by setting $\lambda_1$=0. 

In a DTT, the description of nonequilibrium hydrodynamic states requires the introduction of a new tensor $\xi^{\alpha\gamma}$ which is symmetric, traceless and vanishes in equilibrium \cite{geroch}. For a conformal fluid, $\xi^{\alpha\gamma}$ must be transverse as well. The DTT provides an equation of motion for $\xi^{\alpha\gamma}$ (see Ref. \cite{nos} for details). For a conformal fluid in $2+1$ dimensions the evolution is given by
\begin{equation}
\begin{split}
\partial_\tau \xi^{i\alpha} &= -\frac{2}{3u^\tau}\xi^{i\alpha}\nabla_\mu u^\mu - \frac{1}{\tau_\pi u^\tau}\xi^{i\alpha} + \frac{1}{\tau_\pi u^\tau} \sigma^{i\alpha} \\
&~ - \frac{\lambda_1}{3\tau_\pi \eta u^\tau}\xi^{<i}_\mu \xi^{\alpha> \mu} - \frac{u^i\xi^\alpha_\mu + u^\alpha \xi^i_\mu}{u^\tau}Du^\mu \\
&~ -\frac{u^j}{u^\tau}\partial_j \xi^{i\alpha} ~.
\end{split}
\label{dxi}
\end{equation}

The shear tensor is calculated from the nonequilibrium tensor $\xi^{\alpha\gamma}$ as follows (see App. \ref{dtt}): 
\begin{equation}
\Pi^{\mu\nu} = \eta \xi^{\mu\nu} -\frac{\lambda_1 \tau_{\pi} T^4}{3\eta} (\xi^{\mu\alpha}\xi^\nu_\alpha - \frac{1}{3}\Delta^{\mu\nu}\xi^{\alpha\gamma}\xi_{\alpha\gamma}) ~.
\end{equation}
The transport coefficients of the BRSS and the DTT are the same because the DTT goes over to BRSS at second-order in velocity gradients. Note however that we are ignoring (possible) higher order corrections to the transport coefficients of the DTT.

As independent variables we choose $(\rho,u^x,u^y,\Pi^{xx},\Pi^{xy},\Pi^{yy})$ for the BRSS and $(\rho,u^x,u^y,\xi^{xx},\xi^{xy},\xi^{yy})$ for the DTT. The $\tau$ component of the velocity follows from normalization, $u^\tau=\sqrt{1+u_x^2+u_y^2}$, while the other nontrivial components of $\Pi^{\mu\nu}$ (and of $\xi^{\mu\nu}$) follow from the transversality and tracelessness conditions. 

In order to solve the hydrodynamic equations, we employ the method described in Ref. \cite{br} (see also Ref. \cite{luzum}). The set of six coupled differential equations is cast into a linear system for the time derivatives of the independent variables. This linear system is solved using a finite difference method which is first-order accurate in the temporal grid spacing and second-order accurate in the spatial grid spacing. To be more precise, the derivatives are replaced by:
\begin{equation}
\begin{split}
\partial_\tau f(x,y,\tau) &= \frac{f(x,y,\tau+\delta\tau)-f(x,y,\tau)}{\delta \tau} ~~ \textrm{and} \\
\partial_x f(x,y,\tau) &= \frac{f(x+\delta x,y,\tau)-f(x-\delta x,y,\tau)}{2\delta x}
\end{split}
\end{equation}
and similarly for the derivative in $y$. 
We have made nontrivial tests on the code such as, for example, use uniform initial data in the $2+1$ numerical code to recover $0+1$ dimensional results (already presented in Ref. \cite{nos}).

%\subsection{Initial conditions, EoS and freeze-out}
\subsection{Initial conditions and transport coefficients}

Solution of the hydrodynamic equations requires initial conditions for the six independent variables. For the initial transverse velocity and shear tensor we use $u^x=u^y=0$, which implies vanishing initial vorticity, and $(\Pi^{xx},\Pi^{xy},\Pi^{yy})=0$ or $(\xi^{xx},\xi^{xy},\xi^{yy})=0$. It has been shown in several works \cite{song08,luzum,luzum2,dus08} that the evolution of the shear tensor $\Pi^{\mu\nu}$ is quite insensitive to the initialization values, the difference being appreciable only at very early times. We have verified that the elliptic flow show very little sensitivity to the initialization of the shear tensor as well.
In what follows, we take the initialization time to be $\tau_0=$ 1 fm/c.  

The initial energy density profile is calculated using a simple Glauber model \cite{glau}, in which for impact parameter $b$ we have 
\begin{equation}
\rho(\tau_0,x,y,b)=C\times \sigma T_A(x+\frac{b}{2},y)T_A(x-\frac{b}{2},y)
\end{equation}
where $\sigma=$ 40 mb, $C$ is a constant chosen such that $\rho(\tau_0,0,0,0)$ corresponds to a given initialization temperature $T_0$ (via the equation of state), and $T_A$ is the nuclear thickness function given by
\begin{equation}
T_A(x,y)=\int_{-\infty}^\infty \delta_A (x,y,z) ~dz ~~.
\end{equation}
The function $\delta_A$ is the Woods-Saxon density distribution for gold nuclei
\begin{equation}
\delta_A(x,y,z)=\frac{\delta_0}{1+\textrm{exp}[(|{\mathbf x}|-R_0)/\chi]}
\end{equation}
with ${\mathbf x}=(x,y,z)$, $R_0=$ 6.4 fm and $\chi=$ 0.54 fm. The parameter $\delta_0$ is chosen such that $\int d^3{\mathbf x}~ \delta_A({\mathbf x})=197$ as appropriate for Au nuclei. We note that in all calculations we use a 7.5 fm $\times$ 7.5 fm transverse plane (we have used a 13 fm $\times$ 13 fm transverse plane and found that there is no significant change in our results).

Unless otherwise stated, we use values for the second-order transport coefficients corresponding to a strongly-coupled $\cal{N}=$ 4 Super-Yang Mills (SYM) plasma \cite{ads,sonhydro,bat,nat}: 
\begin{equation}
\tau_\pi = 2(2-\ln 2)\frac{\eta}{sT} ~~~\textrm{and} ~~~ \lambda_1=\frac{\eta}{2\pi T} ~,
\label{sym}
\end{equation}
where $s$ is the entropy density. We will show in Appendix \ref{dep} that our results depend only weakly on the precise value of second-order transport coefficients.

\subsection{Equation of State}
% EOS
The set of hydrodynamic equations must be closed with an equation of state (EoS). Since we are interested in computing elliptic flow of the produced particles we must use an EoS including hadronization. We employ the EoS by Laine and Schr\"{o}der \cite{laine} which connects a high-order weak-coupling perturbative QCD calculation at high temperatures to a hadron resonance gas at low temperatures, via an analytic crossover (as suggested by Lattice QCD calculations \cite{huovani,aoki}). We notice that this EoS is the same as that used in Ref. \cite{luzum}, so that the comparisons we make are meaninful (see also Ref. \cite{huovani}).

\subsection{Freeze-out}
In order to compute the elliptic flow of produced particles, the freeze-out process must be simulated. To do so, we use the corresponding modules of the UVH2+1 code which is described in detail in Ref. \cite{luzum,luzum2}. For completeness, we give here a brief overview of the Cooper-Frye freeze-out prescription \cite{coop} implemented in UVH2+1. For the isothermal freeze-out we use here, the conversion from hydrodynamic to particle degrees of freedom takes place in a three-dimensional hypersurface. The spectrum for a single on-shell particle with momentum $p^\mu=(E,\vec{p})$ and degeneracy $d$ is 
\begin{equation}
E\frac{dN}{d^3 p}=\frac{d}{(2\pi)^3}\int p_\mu ~d\Sigma^\mu~ f (x^\mu,p^\mu)
\label{spec}
\end{equation}
where $d\Sigma^\mu$ is the normal vector on the hypersurface, and $f$ is the {\it non-equilibrium} distribution function, customarily given by Grad's ansatz (see \cite{libro,kinbooks,kin,luzum,luzum2} for details)
\begin{equation}
\begin{split}
f(x^\mu,p^\mu)&= f_0(x^\mu,p^\mu)+\delta f (x^\mu,p^\mu) \\
&= f_0(x^\mu,p^\mu)+f_0(x^\mu,p^\mu)\bigg[1\mp f_0(x^\mu,p^\mu)\bigg]\frac{p_\mu p_\nu \Pi^{\mu\nu}}{2T^2(p+\rho)} \\
&\simeq  \textrm{exp}(\frac{-p_\mu u^\mu}{T})\bigg[1+\frac{p_\mu p_\nu \Pi^{\mu\nu}}{2T^2(p+\rho)} \bigg] ~. 
\end{split}
\label{fneq}
\end{equation}
The approximation in the second line holds when $p>>T$, and it is used in our simulations. It has been shown in Ref. \cite{luzum,luzum2} that the systematic error of this approximation is very small at low tranverse momentum $p_T \lesssim$ 2.5 GeV, so we do not expect our results to have a significant error coming from this approximation. 

The UVH2+1 freeze-out module calculates the spectra for particle resonances with masses up to 2 GeV and then determines the spectra of stable particles including feed-down contributions \cite{reson}. For this last step it uses the AZHYDRO package \cite{AZ}. In this paper we will focus on the minimum-bias elliptic flow coefficient $v_2$ at central rapidity, which is given by 
\begin{equation}
v_2(p_T)=\frac{\int db ~b ~v_0(p_T,b)\tilde{v}_2(p_T,b)}{\int db ~b ~v_0(p_T,b)}
\label{v2min}
\end{equation}
where the coeffients $v_0$ and $\tilde{v}_2$ are related to the particle spectra (including feed-down contributions) by
\begin{equation}
E \frac{d N}{d^3\vec{p}} = v_0(p_T,b)[1+2\tilde{v}_2(p_T,b)\cos(2\phi)]
\label{v0v2}
\end{equation}
with $\phi=\textrm{arctan}(p_y/p_x)$ and $p_T=(p_x^2+p_y^2)^{1/2}$. We note that kinetic and chemical freeze-out occur at the same temperature, which represents a simplification of the real process (see for instance Ref. \cite{chem} and references therein). 

We may obtain an internal consistency check of the hydrodynamical
approximation by computing elliptic flow (\ref{v0v2}) with the left hand side given
by Eq. (\ref{spec}) with the distribution function (\ref{fneq}), or else with only the
equilibrium distribution function. The difference between these two results
is the so-called nonequilibrium contribution to elliptic flow $\delta v_2$. 

It has been shown by Song and Heinz \cite{song08}, by Chaudhuri \cite{ch2,chaud} and by Dusling and Teaney \cite{dus08} that beyond $p_T \sim $2--3 GeV the nonequilibrium correction to the momentum distribution function  becomes comparable to the equilibrium contribution, thus rendering Grad's ansatz unreliable. A way to estimate the value of $p_T$ at which this happens is to compare  $\delta v_2$ to the total $v_2$. 
We note that this way of determining when does Grad's ansatz become unreliable may miss corrections to the particle distribution function which are independent of azimuthal angle. However, it still provides an estimate which gives an idea of how large are the nonequilibrium corrections to the distribution function.

\begin{figure}[htb]
\scalebox{1}{\includegraphics{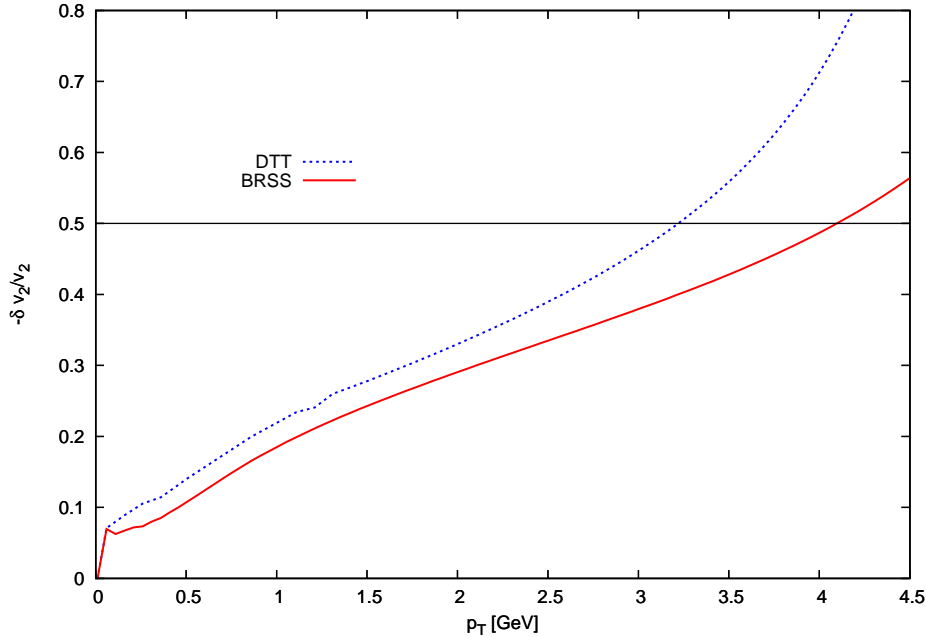}}
\vspace{1cm}
\caption{(Color online) Nonequilibrium contribution to the total elliptic flow $-\delta v_2/v_2$ calculated with the BRSS and DTT with $\eta/s=$ 0.08 and $b=$ 7 fm. The initial and freeze-out temperatures are $T_i=$ 333 MeV and $T_f=$ 140 MeV in both models. The horizontal line indicates the (estimated) breakdown of Grad's ansatz.}
\label{figneq}
\end{figure}

In Figure \ref{figneq} we show $\delta v_2/v_2$ as a function of tranverse momentum, calculated with the BRSS and the DTT with $\eta/s=$ 0.08 and $b=$ 7 fm. In this calculation, we set $T_i=$ 333 MeV and $T_f=$ 140 MeV in both models.

It is seen that, both in the BRSS and in the DTT, the nonequilibrium contribution to $v_2$ is significant even at low $p_T$. In the BRSS $\delta v_2$ is always smaller that in the DTT, indicating that in the latter dissipative corrections to the nonequilibrium distribution function are larger. The difference between $\delta v_2/v_2$ calculated with the DTT and the BRSS grows with $p_T$, in particular for $p_T >$ 3 GeV where $\delta v_2/v_2$ in the DTT starts growing faster. We can estimate the value of $p_T$ at which Grad's ansatz  becomes unreliable as that at which $-\delta v_2/v_2 \sim$ 0.5, which is indicated by a horizontal line in the figure. In the DTT the value of $p_T$ at which Grad's ansatz becomes unreliable turns out to be $p_T \sim$ 3.3 GeV instead of $p_T=$ 4 GeV as inferred in the BRSS.

\section{Results}
\label{res}

In this section we go over to our main goal in this paper, namely to compare the results obtained with the DTT to RHIC data and to the BRSS. We start by comparing the entropy production and the spatial and momentum anisotropies in both hydrodynamic models, and then proceed to compare particle multiplicity, $<p_T>$ and elliptic flow to data in order to constrain the values of $\eta/s$. 

\begin{figure}[htb]
\scalebox{1}{\includegraphics{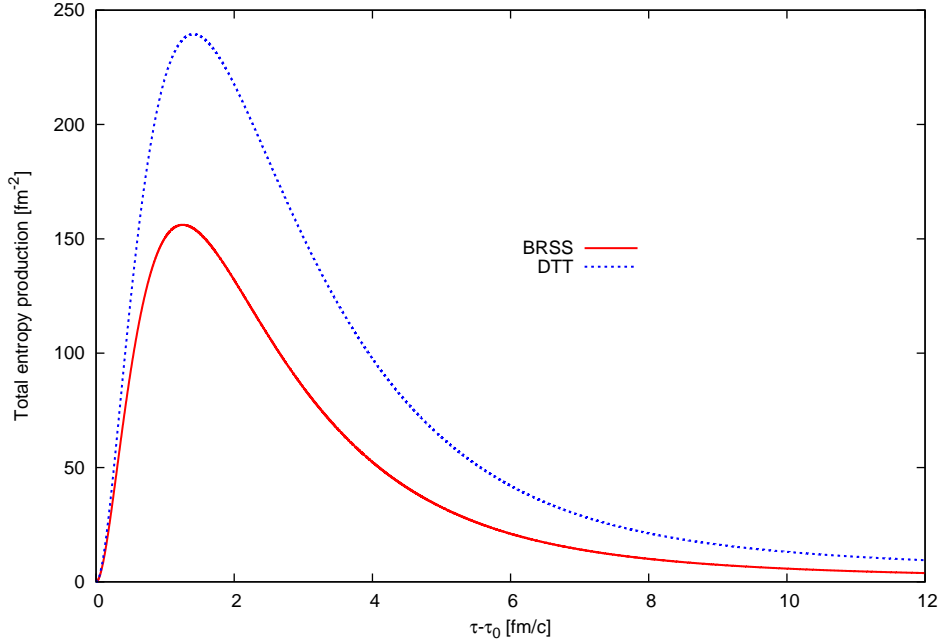}}
\vspace{1cm}
\caption{(Color online) Total entropy production $\sigma$ as a function of $\tau-\tau_0$ for the BRSS and the DTT with $\eta/s=$ 0.08, $T_i=$ 333 MeV, $T_f=$ 140 MeV and $b=7$ fm.}
\label{figent}
\end{figure}

Figure \ref{figent} shows the evolution of the total entropy production $\sigma$ in a collision with impact parameter $b=$ 7 fm, for the DTT and the BRSS with $\eta/s=$ 0.08, an initial temperature $T_i=$ 333 MeV and a freeze-out temperature $T_f=$ 140 MeV. The total entropy production is given by the following expressions 
\begin{equation}
\begin{split}
\sigma_{BRSS} &\equiv \int dx dy ~ (\partial_\mu S^\mu) |_{BRSS} = \int dx dy ~ \frac{\Pi^{\mu\nu}\Pi_{\mu\nu}}{2\eta T} ~~~~ \textrm{and} \\
\sigma_{DTT} &\equiv \int dx dy ~ (\partial_\mu S^\mu) |_{DTT}  = \int dx dy ~ \frac{\eta \xi^{\mu\nu}\xi_{\mu\nu}}{2T} ~~ .
\end{split}
\end{equation}
We note that in order for the comparison to make sense we have used the same $T_i$ and $\eta/s$ for the DTT and the BRSS. It is seen that the entropy production is significantly larger in the DTT. At early (late) times the DTT produces $\sim$ 25 $\%$ ($\sim$ 40 $\%$) more entropy than the BRSS. However, at very early times $\tau - \tau_0 < $ 1 fm/c the entropy production is the same in both models.

\begin{figure}[htb]
\scalebox{1}{\includegraphics{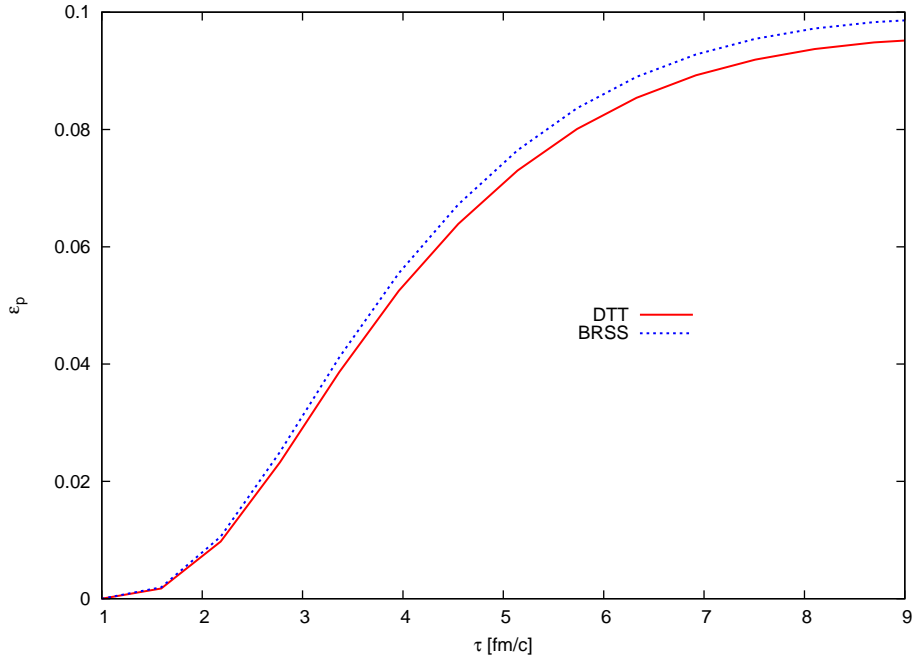}}
\vspace{1cm}
\caption{(Color online) Evolution of the momentum anisotropy for the BRSS and the DTT with $b=7$ fm, $T_i=$ 333 MeV, $T_f=$ 140 MeV and $\eta/s=$ 0.08.}
\label{figepsame}
\end{figure}

We now go over to compare the spatial ($\epsilon_x$) and momentum ($\epsilon_p$) anisotropies, defined as
\begin{equation}
\epsilon_x = \frac{<y^2-x^2>_\rho}{<y^2+x^2>_\rho} ~~ \textrm{and} ~~ \epsilon_p = \frac{<T^{xx}-T^{yy}>}{<T^{xx}+T^{yy}>} ~,
\end{equation}
where $<\ldots>_\rho$ denotes an average procedure over the tranverse plane with the energy density $\rho$ as weighting factor. 
In Figure \ref{figepsame} we show $\epsilon_p$ calculated with the same value of $\eta/s=$ 0.08, $T_i=$333 MeV and $T_f=$ 140 MeV for the BRSS and the DTT. This allows us to determine the differences between both hydrodynamic formalisms. It is seen that $\epsilon_p$ is slightly smaller in the DTT, indicating that the DTT leads to larger shear stress and thus to larger dissipation. This is consistent with the fact that, as already discussed, more entropy is produced in the DTT.

\begin{figure}[htb]
\scalebox{1}{\includegraphics{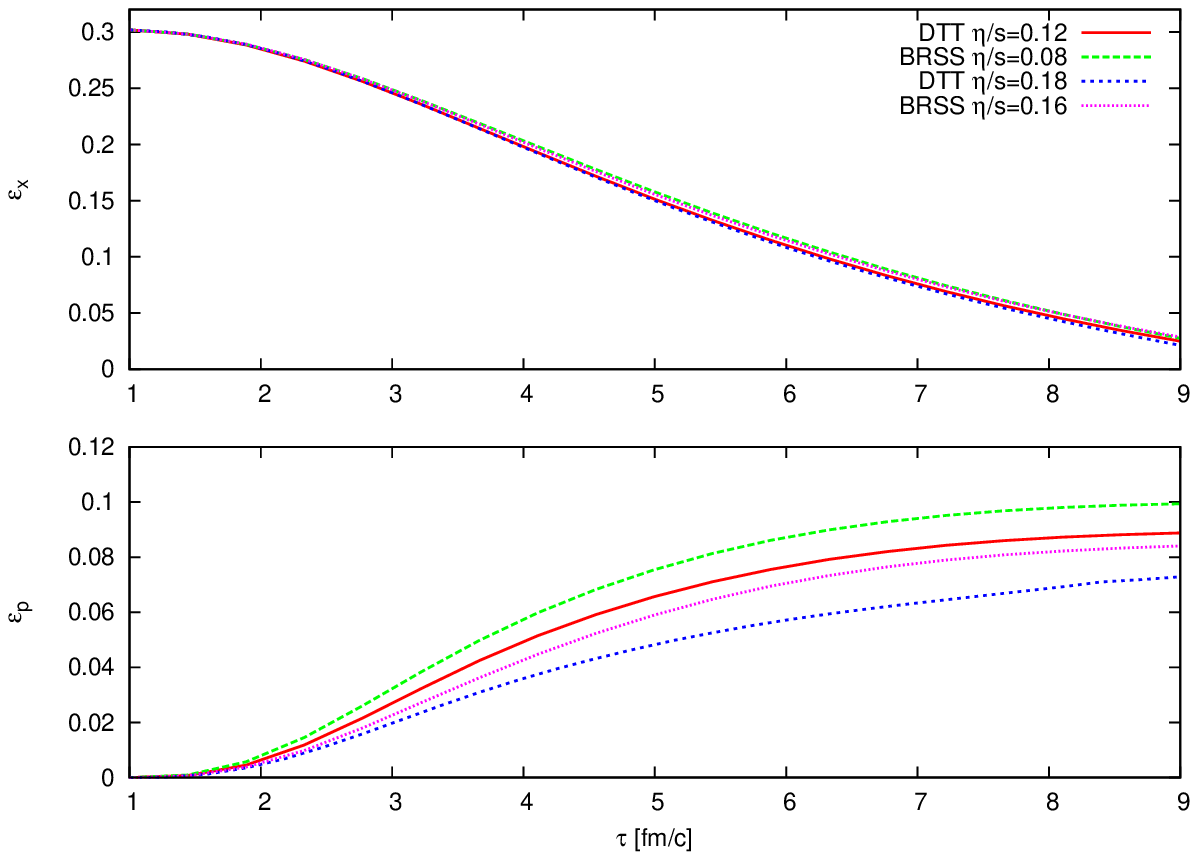}}
\vspace{1cm}
\caption{(Color online) Evolution of the spatial (upper panel) and momentum (lower panel) anisotropies for the BRSS and the DTT with $b=7$ fm, $T_i=$ 333 MeV, $T_f=$ 140 MeV and different values of $\eta/s$.}
\label{figep}
\end{figure}

We now allow the values of $\eta/s$ to vary. In Figure \ref{figep} we show the evolution of $\epsilon_x$ and $\epsilon_p$  for the BRSS and the DTT with $b=$ 7 fm, $T_i=$333 MeV, $T_f=$ 140 MeV and several values of $\eta/s$. The values of $\eta/s$ correspond to those that will be suitable for matching Kaon multiplicity and $<p_T>$ to data, as will be shown later.

It is seen that the spatial anisotropies calculated in the DTT and in the BRSS are very similar, and quite independent of $\eta/s$. In contrast, the momentum anisotropies are seen to strongly depend on $\eta/s$ and are very different in the BRSS and the DTT. The DTT $\epsilon_p$ is systematically lower that the BRSS one, and, as we will see, this will be reflected on the calculated elliptic flow. 

In order to compare the elliptic flow calculated with the DTT or the BRSS to data and in this way constrain the value of $\eta/s$, we will follow the procedure of Luzum and Romatschke in \cite{luzum}. The idea is to determine the initial temperature $T_i$ for $b=$ 0 and $T_f$ by matching the hydrodynamic simulation to total multiplicity and $<p_T>$. 
As explained in Ref. \cite{luzum}, one should refrain from trying to match pion multiplicity and $<p_T>$ since the Boltzmann's approximation used in Eq. (\ref{fneq}) leads to an unavoidable systematic error. We will therefore aim at a reasonable fit of Kaon multiplicity and $<p_T>$ to determine $T_i$ and $T_f$. In Appendix \ref{dep} we will show that the results for elliptic flow are only weakly dependent on the values of second-order transport coefficients $(\tau_{\pi},\lambda_1)$. For this reason, in the remaining of this section we will use coefficients appropiate for a SYM plasma as given in Eq. (\ref{sym}).

We start by presenting the results for total Kaon multiplicity and $<p_T>$ obtained in both models with $\eta/s=$ 0.08, $T_i=$ 333 MeV and $T_f=$ 140 MeV, compared to PHENIX data \cite{phenix}. As in the case of spatial and momentum anisotropies, this allows us to evaluate the differences in both hydrodynamic formalisms for the same values of $\eta/s$, $T_i$ and $T_f$. We also show the results for the DTT with $\eta/s=$ 0.12 and $T_i=$ 328 MeV (which as shown later is the value of $T_i$ that gives, for this value of $\eta/s$ in the DTT, the best matching to data).

\begin{figure}[htb]
\scalebox{1}{\includegraphics{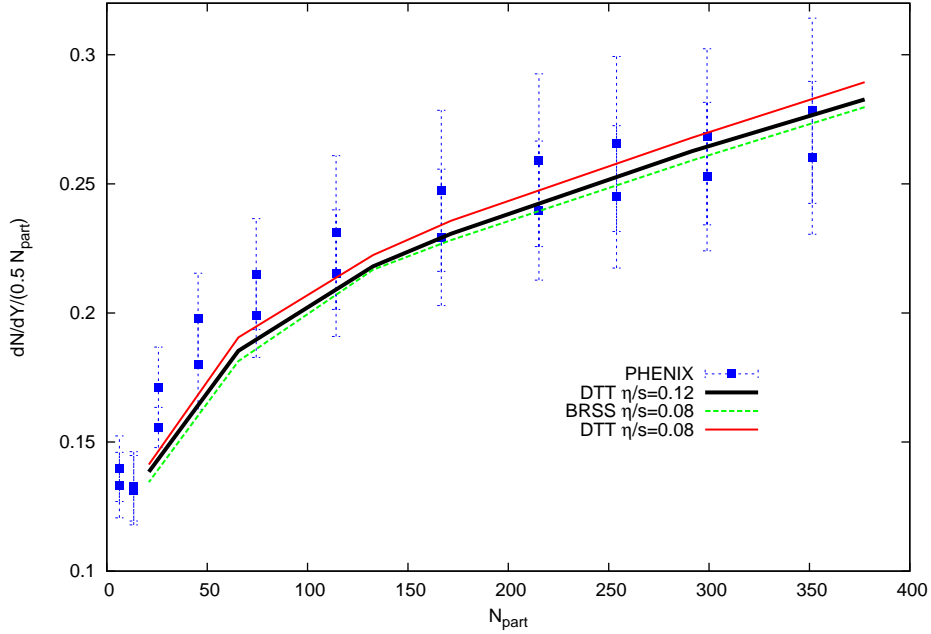}}
\vspace{1cm}
\caption{(Color online) Centrality dependence of total multiplicity for Kaons for Au+Au collisions at $\sqrt{s}=$ 200 GeV compared to the DTT and the BRSS for $\eta/s=$0.08, $T_i=$ 333 MeV and $T_f=$ 140 MeV. We also show the DTT result with $\eta/s=$ 0.12 and $T_i=$ 328 MeV for comparison. Data is from the PHENIX Collaboration \cite{phenix}. The two sets of datapoints correspond to $K^+$ and $K^-$.}
\label{msame}
\end{figure}

\begin{figure}[htb]
\scalebox{1}{\includegraphics{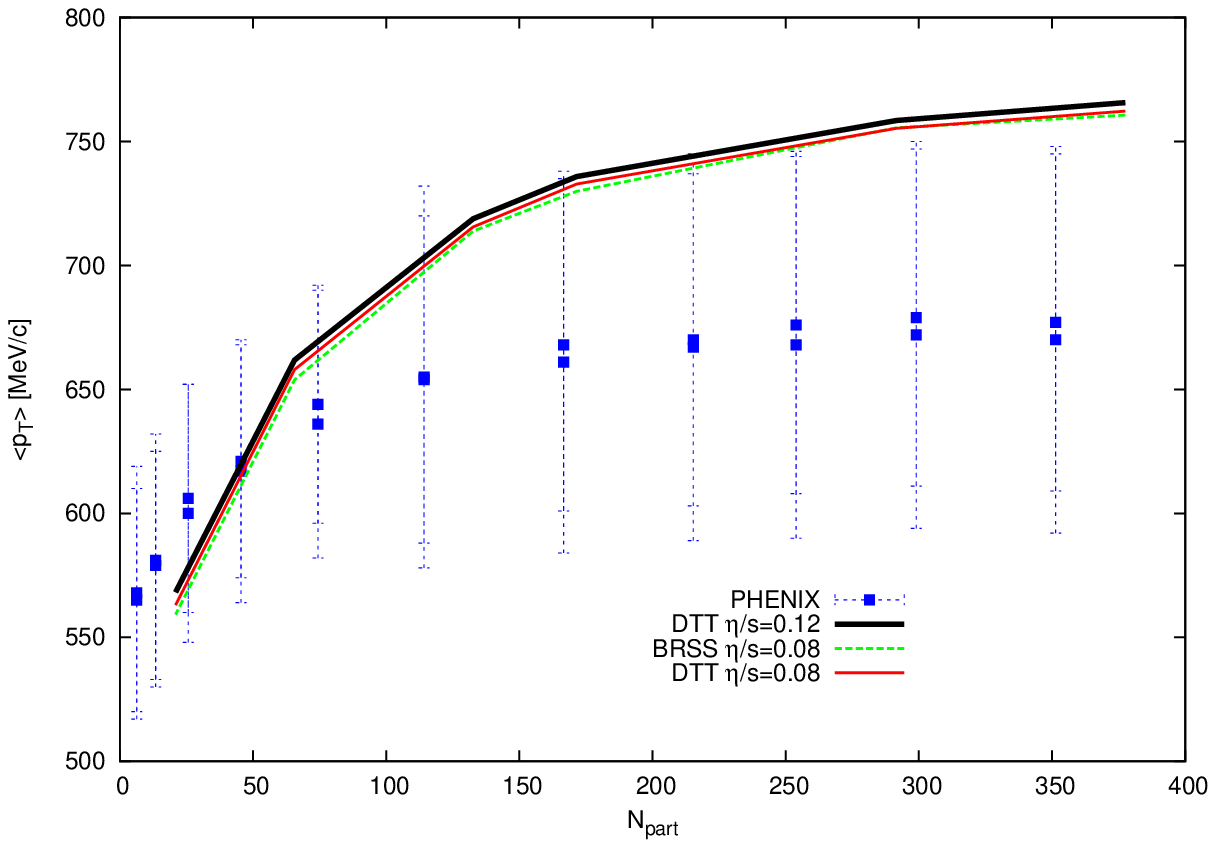}}
\vspace{1cm}
\caption{(Color online) Centrality dependence of $<p_T>$ for Kaons for Au+Au collisions at $\sqrt{s}=$ 200 GeV compared to the DTT and the BRSS for $\eta/s=$0.08, $T_i=$ 333 MeV and $T_f=$ 140 MeV. We also show the DTT result with $\eta/s=$ 0.12 and $T_i=$ 328 MeV for comparison. Data is from the PHENIX Collaboration \cite{phenix}.}
\label{ptsame}
\end{figure}

Figure \ref{msame} shows the total multiplicity. The two sets of datapoints correspond to $K^+$ and $K^-$, which can not be distinguished in our model since the EoS corresponds to zero net-baryon density. It is seen that, in order to achieve the same multiplicity as in the BRSS, we must take $\eta/s=$ 0.12 in the DTT. Figure \ref{ptsame} shows $<p_T>$ obtained in both models. It is seen that $<p_T>$ is larger in the DTT than in the BRSS, indicating that the DTT leads to larger transverse flow. Larger shear stress leads to larger transverse flow, so the results obtained for $<p_T>$ confirm the fact that the DTT leads to stronger dissipation and consequently, larger entropy production.

\begin{figure}[htb]
\scalebox{1}{\includegraphics{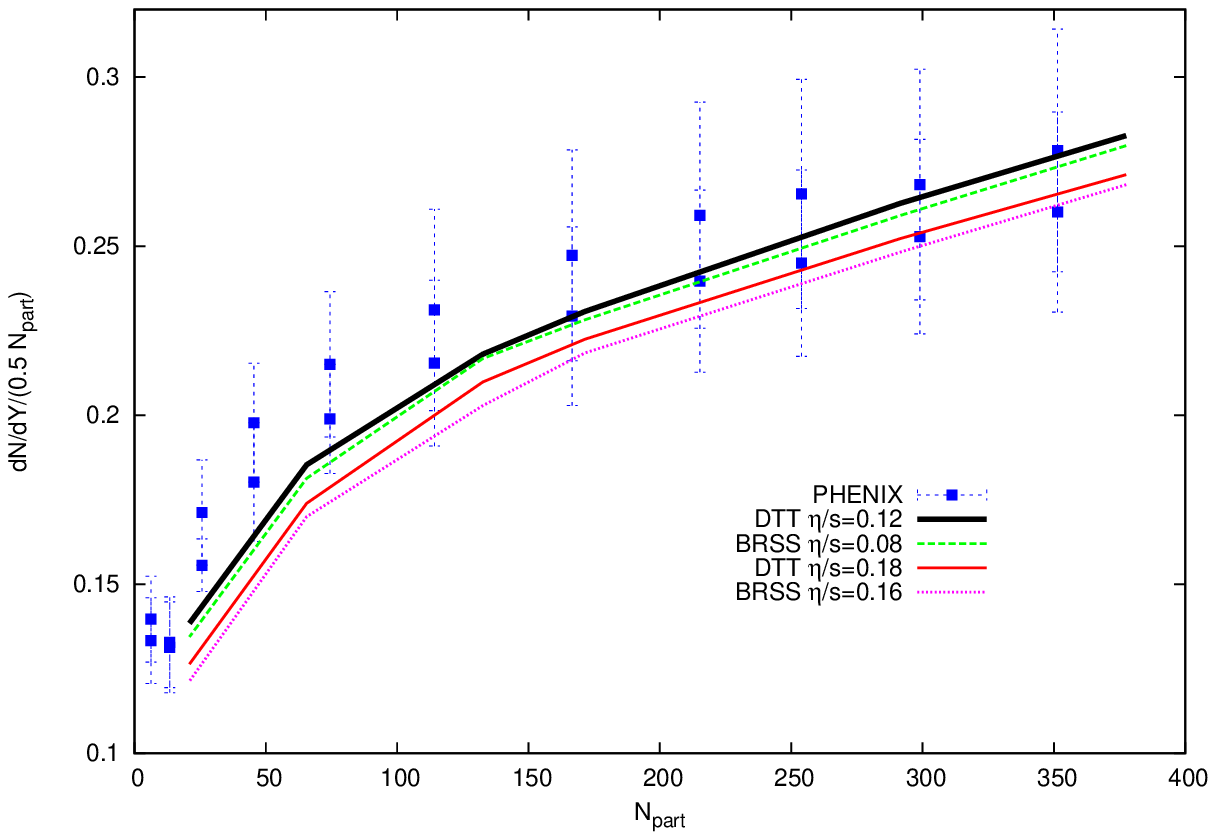}}
\vspace{1cm}
\caption{(Color online) Centrality dependence of total multiplicity for Kaons for Au+Au collisions at $\sqrt{s}=$ 200 GeV compared to the DTT and the BRSS for various values of $\eta/s$. The freeze-out temperature is $T_f=$ 140 MeV. Data is from the PHENIX Collaboration \cite{phenix}.}
\label{figm}
\end{figure}

\begin{figure}[htb]
\scalebox{1}{\includegraphics{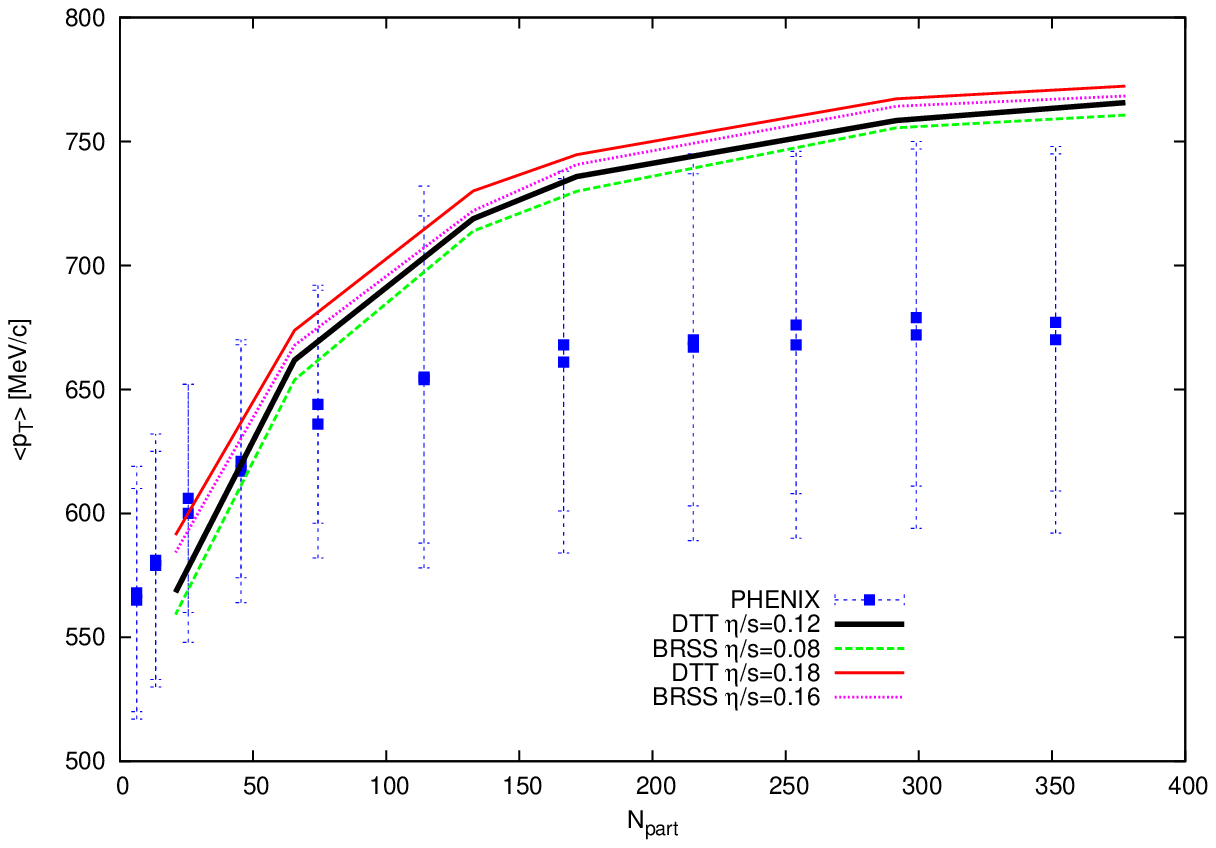}}
\vspace{1cm}
\caption{(Color online) Centrality dependence of $<p_T>$ for Kaons for Au+Au collisions at $\sqrt{s}=$ 200 GeV compared to the DTT and the BRSS for various values of $\eta/s$. The freeze-out temperature is $T_f=$ 140 MeV. Data is from PHENIX Collaboration \cite{phenix}.}
\label{figp}
\end{figure}

We now proceed to find $T_i$ by matching total Kaon multiplicity and $<p_T>$ to data. We note that we fix $T_f=$ 140 MeV in both models. Figures \ref{figm} and \ref{figp} show the total multiplicity and $<p_T>$ calculated in the BRSS and the DTT. By performing this combined matching to multiplicity and $<p_T>$ data, we find for the DTT that $T_i=$ 328 MeV (323 MeV) at $\eta/s=$0.12 (0.18) gives a reasonable fit to data, comparable to that obtained by the BRSS with $T_i=$ 333 MeV (327 MeV) at $\eta/s=$0.08 (0.16) \cite{luzum}. 

\begin{figure}[htb]
\scalebox{1}{\includegraphics{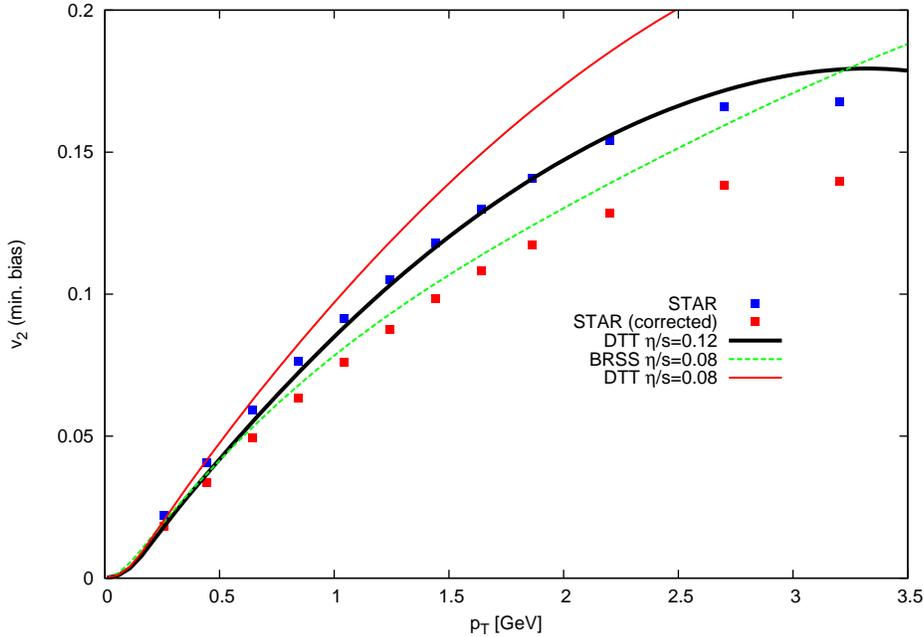}}
\vspace{1cm}
\caption{(Color online) Comparison of STAR data on charged-hadron minimum-bias elliptic flow to DTT and BRSS results with $\eta/s=$ 0.08 and $T_i=$ 333 MeV. We also show results of the DTT with $\eta/s=$ 0.12 and $T_i=$ 328 MeV for comparison.}
\label{v2same}
\end{figure}
\begin{figure}[htb]
\scalebox{1}{\includegraphics{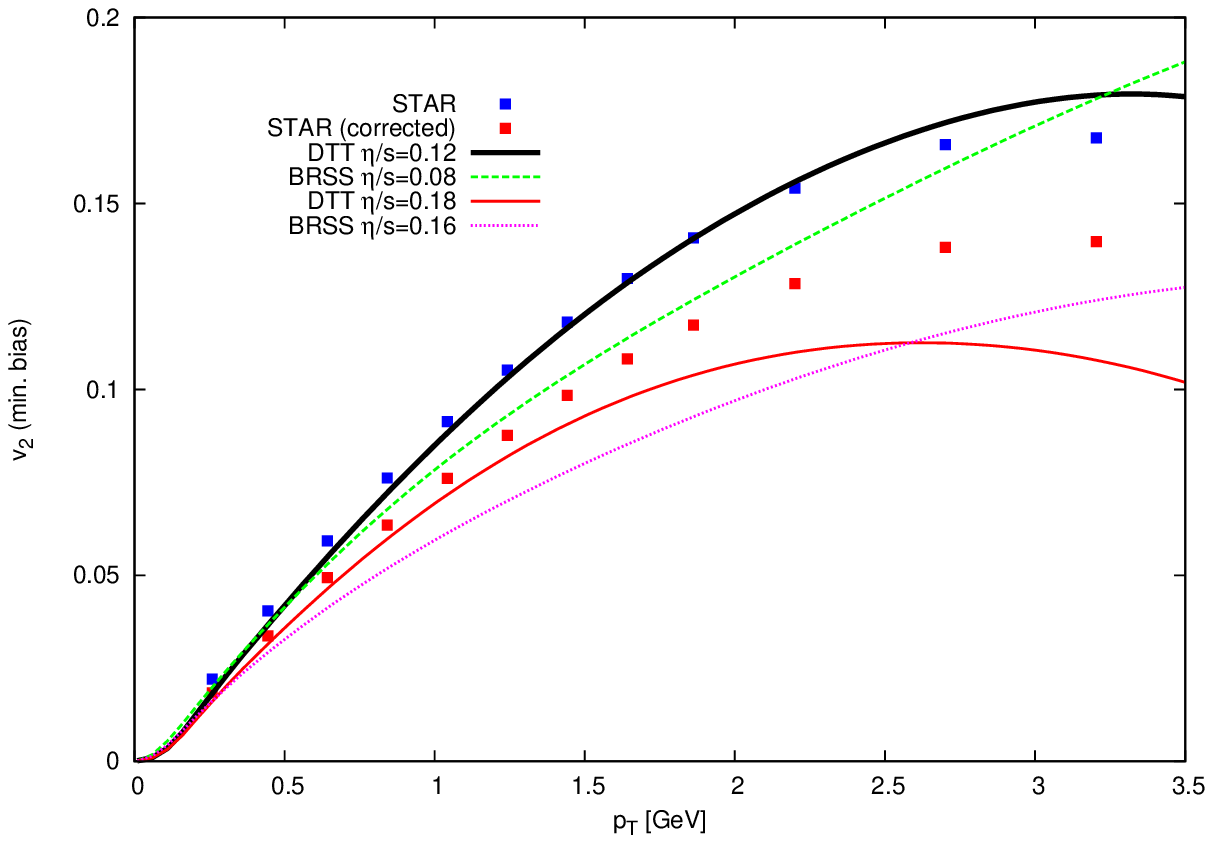}}
\vspace{1cm}
\caption{(Color online) Comparison of DTT results and experimental data on charged-hadron minumum-bias elliptic flow by STAR (event-plane) \cite{star}. STAR event-plane data has been reduced by 20 $\%$ (red squares) to account for non-flow contributions (estimatively) \cite{star}. We also show results for the BRSS for comparison.}
\label{figv}
\end{figure}

With the values for $T_i$ and $T_f$ obtained before, we now go over to calculate the charged-hadron minimum-bias elliptic flow with the DTT and the BRSS and compare the results to data. 
In Figure \ref{v2same} we show the elliptic flow calculated in the DTT and the BRSS with $\eta/s=$ 0.08 and $T_i=$ 333 MeV, and in the DTT with $\eta/s=$ 0.12 and $T_i=$ 328 MeV, compared to experimental data from the STAR Collaboration \cite{star}. We note that in order to estimate the removal of nonflow contributions to the elliptic flow, we reduce the STAR data by 20$\%$ (see Refs. \cite{star,luzum}). 
 It is seen that using the DTT with $\eta/s=$ 0.08 provides a poor fit to data, specially at intermediate and large transverse momentum. 

Figure \ref{figv} shows the charged-hadron minimum-bias elliptic of the BRSS and the DTT for different values of $\eta/s$, compared to data. The most important differences in the calculated $v_2$ with the BRSS and with the DTT take place for $p_T>$ 2 GeV. It is seen that $v_2$ reaches its maximum at lower values of $p_T$ in the DTT, and that the values for $\eta/s$ for which the DTT model is consistent with data ($\eta/s \le$ 0.18) are slightly larger than those for the BRSS. The fact that the values of $\eta/s$ for which the elliptic flow is consistent with data are the same as those obtained from matching Kaon multiplicity and $<p_T>$ does not mean that $v_2$ measurements do not constrain $\eta/s$, but rather that viscous hydrodynamics provides a consistent description of these observables in Au+Au collisions.

We now wish to constrain the value of $\eta/s$ and give an upper bound beyond which our results cease to be consistent with experimental data.  
From Figures \ref{msame}-\ref{figv} we find that we can match the DTT results to experimental data provided $\eta/s \le$ 0.18. It is difficult to determine the uncertainty in $\eta/s$ coming from the hydrodynamic simulation. Considering the dependence of final results on the following factors (i) the precise value of second-order transport coefficients (see Appendix \ref{dep}), (ii) the mesh grid (see Appendix \ref{dep}) and (iii) the precise value of $T_i$ and $T_f$ obtained from matching hydrodynamic results to data on Kaon multiplicity and $<p_T>$, as sources of uncertainty in the determination of $\eta/s$, we can estimate this theoretical uncertainty in $\pm$ 0.07. Taking into account the uncertainty in the removal of non-flow contributions to the measured charged-hadron $v_2$ (min. bias) by STAR \cite{star}, we can, following Ref. \cite{luzum}, estimate the experimental uncertainty in $\pm$ 0.1. Therefore, we conclude that the DTT model favors $\eta/s \le 0.35$. We emphasize that our estimate for $\eta/s$ does not contemplate the uncertainty coming from several other factors such as bulk viscosity \cite{koide,koidebul,monnai}, different temperatures for kinetic and chemical freeze-out \cite{chem}, precise knowledge of the EoS and of the initial conditions \cite{luzum,luzum2,song08,huovani}, which are expected to have a significant influence on the value of $\eta/s$. For this reason, our estimate for $\eta/s$ should be regarded as a conservative one. One should note, however, that both the DTT and the BRSS are consistent with experimental data, although none of the models can reproduce the saturation of the  measured minimum bias elliptic flow. Recent studies \cite{rad,luzgrad,monnai,koidebul} suggest that the origin of this failure of viscous hydrodynamics to reproduce saturation of the ellipic flow is Grad's quadratic ansatz for the nonequilibrium correction to the thermal distribution function given by Eq. (\ref{fneq}). 
  
Finally, we note that this result for $\eta/s$ is in good agreement with the upper bound found in several other works \cite{luzum,luzum2,bound} by similar matching of viscous hydrodynamics to data, and supports the notion that the matter created at RHIC exhibits almost perfect fluidity.

\section{Summary and conclusions}
\label{concsec}

We have studied the space-time evolution of a conformal plasma in $2+1$ dimensions using second-order as well as divergence-type dissipative hydrodynamics. In the simulations, we employed a simple Glauber model to calculate the initial energy density distribution, a model equation of state with an analytic crossover, and the Cooper-Frye prescription for isothermal freeze-out (the latter implemented in the code UVH2+1 \cite{luzum,romyrom}).

We have made a comparison of the calculated Kaon total multiplicity and $<p_T>$ with experimental data by the PHENIX Collaboration \cite{phenix}, as well as of the elliptic flow with experimental data by the STAR Collaboration \cite{star}. We have found that the difference between the BRSS and the DTT elliptic flows starts to become significant when $p_T>$ 2 GeV: the elliptic flow calculated with the DTT reaches its maximum at lower values of $p_T$. Including an estimate for the uncertainty in the determination of $\eta/s$ from data, we find that the DTT can be matched to RHIC data provided $\eta/s \le 0.35$, in good agreement with previous studies based on Israel-Stewart or BRSS equations. The results we obtain also show that the differences between hydrodynamic formalisms are a significant source of uncertainty in the precise extraction of $\eta/s$ from data.
 
We note that niether the DTT nor the BRSS are able to reproduce the experimental saturation of elliptic flow at $p_T \gtrsim$ 2.5 GeV, possibly pointing to the incorrectness of Grad's ansatz for the nonequilibrium distribution function (Eq. (\ref{fneq})). 
A related aspect that surely deserves further investigation is the inclusion of bulk viscosity in the hadronic stage of the fireball's evolution and during the deconfinement crossover. In this respect, it should be noted that recent work \cite{koidebul} has raised concern about the validity of Grad's ansatz for a reliable computation of freeze-out when bulk viscosity is present (see also Ref. \cite{monnai} for a different treatment of bulk viscosity). It was found in Ref. \cite{koidebul} that the nonequilibrium corrections due to bulk viscosity to the distribution function are considerably larger than those coming from shear viscosity, rendering the application of Grad's moment method doubtful when bulk viscosity is taken into account. Moreover, Luzum and Ollitrault \cite{luzgrad} have found that even for a conformal plasma Grad's ansatz is disfavored by data on $v_4/(v_2)^2$, while Dusling, Moore and Teaney
\cite{rad} have calculated the momentum dependence of the nonequilibrium contribution to the distribution function in a weak coupling setting and found it proportional to $p_T^{3/2}$.  

It would be therefore interesting to study this point within the framework of DTTs. In order to further investigate this issue, the relation between the DTT and microscopic theory must be precisely determined.

\begin{acknowledgments}
We are grateful to Paul Romatschke, Tomoi Koide and Akihiko Monnai for useful comments and interesting discussions. We thank Roy Lacey for bringing Refs. \cite{lac} to our attention, and A. Poskanzer for providing data from the STAR Collaboration. This work has been supported in part by ANPCyT, CONICET and UBA (Argentina).
\end{acknowledgments}

\appendix
\section{Dependence on transport coefficients and grid}
\label{dep}

In this Appendix we will evaluate the dependence of the results obtained with the DTT on the spatial mesh and on the values for second-order transport coefficients. The parameters used for this evaluation are $\eta/s=$ 0.08, $\tau_0=$ 1 fm/c, $T_i=$ 333 MeV and $T_f=$ 140 MeV in both cases.

\begin{figure}[htb]
\scalebox{1}{\includegraphics{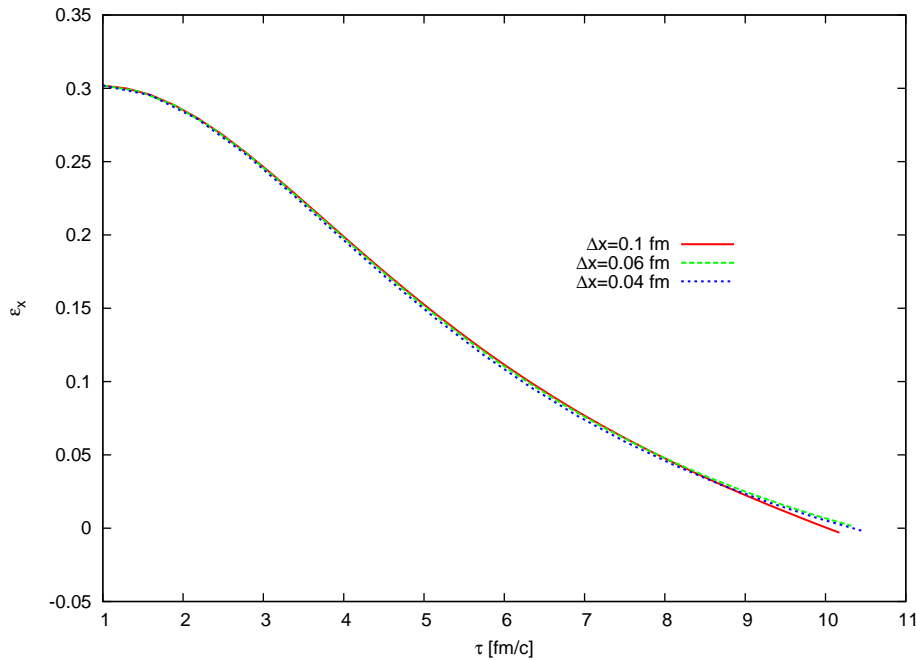}}
\vspace{1cm}
\caption{(Color online) Evolution of the spatial anisotropy in the DTT with $\eta/s=$ 0.08 and $b=$ 7 fm, for different values of the space grid $\Delta x$.}
\label{figmesh}
\end{figure}

In Figure \ref{figmesh} we show the evolution of spatial anisotropy $\epsilon_x$ for a collision with $b=$ 7 fm and a time step $\Delta \tau =$ 0.002 fm/c, for different values of the space grid $\Delta x =$ 0.1, 0.06 and 0.04 fm. It is seen that the dependence of $\epsilon_x$ on $\Delta x$ is very small and only significant at late times. There is practically no difference between the results obtained with $\Delta x=$ 0.06 fm and 0.04 fm, and for this reason we employ $\Delta x =$ 0.06 fm in all the simulations shown in this work. We have also checked that diminishing the time step did not change the results appreciably, and therefore use $\Delta \tau =$ 0.002 fm/c throughout.

\begin{figure}[htb]
\scalebox{1}{\includegraphics{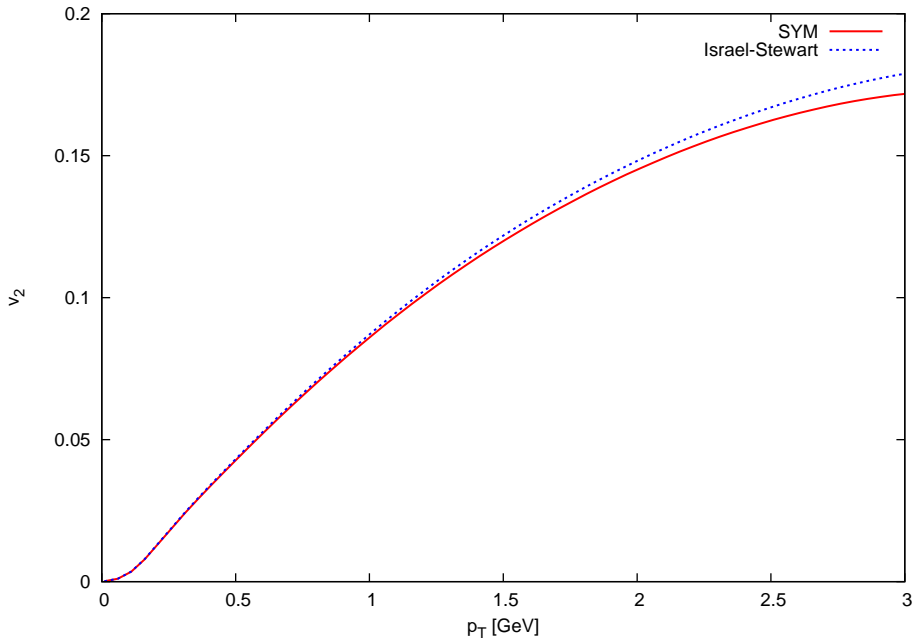}}
\vspace{1cm}
\caption{(Color online) Charged-hadron elliptic flow calculated in the DTT with $\eta/s=$ 0.08 and $b=$ 7 fm with $\tau_{\pi}=2(2-\ln 2)\eta/sT$ and $\lambda_1=\eta /2\pi T$, corresponding to a SYM plasma, and with $\tau_{\pi}=6\eta /sT$ and $\lambda_1=$ 0, corresponding to weakly-coupled Israel-Stewart theory.}
\label{figt}
\end{figure}

A convenient quantity to measure the influence of second-order coefficients on final hadron observables is the elliptic flow coefficient $\tilde{v}_2$ (see Eq. (\ref{v0v2})). In Figure \ref{figt} we show the charged-hadron elliptic flow calculated in the DTT with $\eta/s=$ 0.08 and $b=$ 7 fm, for two sets of second-order coefficients, namely those of a SYM plasma given in Eq. (\ref{sym}) and those corresponding to weakly-coupled Israel-Stewart formalism ($\tau_{\pi}=6\eta/sT$ and $\lambda_1=$ 0). It is seen that the difference between $v_2$ calculated with both sets of coefficients is negligable for $p_T <$ 1.5 GeV and small for higher $p_T$. It is interesting to note that, although both curves are very similar, the elliptic flow calculated for the SYM plasma reaches its maximum at a lower value of $p_T$. We conclude that the results depend only weakly on the precise values of $\tau_{\pi}$ and $\lambda_1$, at least for low values of $\eta/s$. This is in agreement with the findings of Ref. \cite{luzum}.

\section{Divergence-type theories}
\label{dtt}

In this Appendix we give a brief summary of divergence-type theories (DTTs). Detailed discussions can be found in Refs. \cite{geroch,calz98,liu}.

According to Geroch and Lindblom \cite{geroch}, the hydrodynamical description of a nonequilibrium state requires, besides the particle current $N_\mu$ and the stress-energy tensor $T_{\mu\nu}$, a new third order tensor $A_{\mu\nu\rho}$ obeying an equation of motion of divergence type. The dynamical equations are the conservation laws of $N_\mu$ and $T_{\mu\nu}$, together with an equation describing the dissipative part:
\begin{equation}
D_{\mu} A^{\mu\nu\rho} = I^{\nu\rho}
\label{eomapp}
\end{equation}
where $A^{\mu\nu\rho}$ and $I^{\nu\rho}$ are algebraic local functions of $N^\mu$ and $T^{\mu\nu}$ and symmetric in the indices $(\nu,\rho)$, and $D_{\mu}$ is the covariant derivative. 
The entropy current is extended to
\begin{equation}
S^\mu = \Phi^\mu - \beta_\nu T^{\mu\nu} - \alpha N^\mu - A^{\mu\nu\rho} \xi_{\nu\rho}
\end{equation}
where $\beta_\nu=u_\nu/T$ is the temperature vector, $\alpha=\mu/T$ is the affinity, $\Phi^\mu$ is the thermodynamic potential and $\xi_{\nu\rho}$ is symmetric, traceless and vanish in equilibrium. 

We now require that the entropy and the thermodynamical potential be algebraic functions of $(\alpha,\beta_\mu,\xi_{\mu\nu})$. If the entropy production is to be nonnegative, then
\begin{equation}
\frac{\partial \Phi^\mu}{\partial \alpha} = N^\mu; \qquad
\frac{\partial \Phi^\mu}{\partial \beta_\nu} = T^{\mu\nu}; \qquad
\frac{\partial \Phi^\mu}{\partial \xi_{\nu\rho}} = A^{\mu\nu\rho}
\label{pphi}
\end{equation}
Thus, as a consequence of the equations of motion, the entropy production rate is 
\begin{equation}
D_\mu S^\mu = -I^{\nu\rho}\xi_{\nu\rho} ~. 
\end{equation}

Since the stress-energy tensor is symmetric, we must also have
\begin{equation}
\Phi^\mu=\frac{\partial \chi}{\partial \beta_\mu}
\end{equation}
where $\chi(\alpha,\beta_\mu,\xi_{\mu\nu})$ is the so-called generating function of the theory. This means that every DTT is completely determined once $\chi$ and $I$ are specified as algebraic functions of $\alpha,\beta_\mu,\xi_{\mu\nu}$. The theory thus constructed satisfies the principles of relativity and entropy, and fully exploits the latter \cite{liu}.

Introducing the symbol $\zeta^A$ to denote the set $(\alpha,\beta_\mu,\xi_{\mu\nu})$, $A^\mu_B$ the set $(N^\mu,T^{\mu\nu},A^{\mu\nu\rho})$ and $I_B$ the set $(0,0,I_{\mu\nu})$, the theory is summed up in the equations
\begin{equation}
\begin{split}
A^\mu_B &= \frac{\partial \Phi^\mu}{\partial \zeta^B} \\
D_\mu S^\mu &= -I_B \zeta^B \\
D_\mu A^\mu_{B} &= I_B ~~.
\end{split}
\end{equation}

We will now review the main results of Ref. \cite{nos} where a quadratic DTT for a conformal fluid in $d=$ 4 dimensions was developed. The most general generating function $\chi$ which is quadratic in $\xi^{\mu\nu}$ is
\begin{equation}
\chi = \chi_0(T) + \chi_1(T)\xi_{\mu\nu}u^\mu u^\nu + \frac{1}{2T^\alpha}(A u_\rho\xi^{\rho\sigma}\xi_{\sigma\tau}u^\tau+B \xi^{\rho\sigma}\xi_{\rho\sigma})
\label{genchi}
\end{equation}
where $(\alpha,A,B)$ are coefficients to be determined.

For a conformal field, $T^{\mu\nu} \rightarrow e^{6 \omega(x^\gamma)} T^{\mu\nu}$ under a Weyl transformation $g_{\mu\nu}\rightarrow e^{-2\omega(x^\gamma)}g_{\mu\nu}$, and $g_{\mu\nu} T^{\mu\nu}=$0. These constraints pose no problem for $\chi_0(T)$ and $\chi_1(T)$, and in combination with the second equation of (\ref{pphi}) lead to 
\begin{equation}
\begin{split}
\chi_0 (T)&= \frac{a}{6}T^2 ~~ , \\
\chi_1(T) &= \frac{\eta}{2} T^{-2} ~~ \textrm{and} \\
\end{split}
\end{equation}
where the energy density is $\rho=aT^4$. 

For the quadratic part of $\chi$, it is necessary to redefine the temperature in order to satisfy both constraints. To ensure that $T^{\mu\nu}$ has the correct conformal weight we need $\alpha=$6. The quadratic stress-energy tensor obtained from $\chi$ then reads
\begin{equation}
T_2^{\mu\nu}=\frac{1}{2}T^{-4}\left[ B(30u^\mu u^\nu-6\Delta^{\mu\nu})\xi^{\rho\sigma}\xi_{\rho\sigma} + 2A \xi^{\mu\rho}\xi^\nu_\rho \right] ~~ .
\end{equation}
To get $T_{2,\mu}^\mu=0$ we need $A=24B$. The idea is to put 
\begin{equation}
T_2^{\mu\nu}=\Pi^{\mu\nu}+\delta T_2^{\mu\nu}
\end{equation}
with $\Pi^{\mu\nu}$ traceless and transverse, and $\delta T_2^{\mu\nu}$ traceless.
We get 
\begin{equation}
\Pi_2^{\mu\nu}=AT^{-4}\left[ \xi^{\mu\rho}\xi^\nu_\rho -\frac{1}{3}\Delta^{\mu\nu}\xi^{\rho\sigma}\xi_{\rho\sigma} \right]
\end{equation}
and 
\begin{equation}
\delta T_2^{\mu\nu}=\frac{5}{8}T^{-4} A (u^\mu u^\nu+\frac{1}{3}\Delta^{\mu\nu})\xi^{\rho\sigma}\xi_{\rho\sigma} ~~ .
\end{equation}
Defining the physical temperature $T_p$ by $2T^4=T_p^4+\sqrt{T_p^8-(5A/2a)\xi^{\rho\sigma}\xi_{\rho\sigma}}$ we obtain 
\begin{equation}
T^{\mu\nu}=\rho_p u^\mu u^\nu + p_p \Delta^{\mu\nu} + \Pi^{\mu\nu}
\end{equation}
with
\begin{equation}
\Pi^{\mu\nu} =\eta\xi^{\mu\nu} +AT^{-4}\bigg(\xi^{\mu\alpha}\xi_\alpha^\nu-\frac{1}{3}\Delta^{\mu\nu}\xi^{\alpha\gamma}\xi_{\alpha\gamma} \bigg) 
\end{equation}
and $\rho_p=aT_p^4$.

The source term is written as $I^{\mu\nu}=I^{\mu\nu}_1+I^{\mu\nu}_2$, where $I^{\mu\nu}_{1,2}$ are linear and quadratic in $\xi^{\rho\sigma}$, respectively. $I^{\mu\nu}_1$ is obtained by requiring the DTT to reproduce Eckart's theory at first order in velocity gradients, while $I^{\mu\nu}_2$ is obtained by requiring that the quadratic DTT satisfy the Second Law exactly. The result is
\begin{equation}
I^{\mu\nu}=-\frac{\eta}{2T} \xi_{\mu\nu}+gT^{-8} \Delta^{\mu\nu}\xi_{\rho\sigma}\xi^{\rho\sigma}
\end{equation}
whereby the entropy production reads 
\begin{equation}
D_\mu S^\mu = \frac{\eta}{2T} \xi_{\rho\sigma}\xi^{\rho\sigma} ~.
\end{equation}

By requiring the DTT to reproduce the BRSS when $\xi^{\mu\nu}$ is expanded at second-order in velocity gradients, the relation between $(A,g)$ and $(\tau_\pi,\lambda_1)$ is found to be
\begin{equation}
 A=-\frac{\lambda_1 \tau_{\pi}}{3\eta}T^8 
\end{equation}
and 
\begin{equation}
g= -\frac{\lambda_1 T^7}{9} ~.
\end{equation}

The equation of motion for the third-order tensor $D_{\mu} A^{\mu\nu\rho} = I^{\nu\rho}$ renders an evolution equation for $\xi^{\nu\rho}$. In $2+1$ the evolution is given by Eq. (\ref{dxi}). 
We note that in Ref. \cite{nos} we have set $T=T_p$ since the extra terms arising when using $T_p$ are of higher order (i.e. are terms which would be obtained from a cubic generating function $\chi$). We have numerically checked that these extra terms are negligible, thus one can use $T=T_p$ without appreciable change in the results.

\end{document}